# Refractive index sensing near exceptional point of a system of triple microcavity

Priyanka Chaudhary[1] and Akhilesh Kumar Mishra[2*]

[1,2]Department of Physics & [2]Centre of Photonics and Quantum Communication Technology, Indian Institute of Technology Roorkee, Roorkee 247667, India.

*E-mail addresses:* pchaudhary@ph.iitr.ac.in[†], akhilesh.mishra@ph.iitr.ac.in[*].

**Abstract**

Degeneracies of non-Hermitian Hamiltonian i.e., exceptional points (EPs) of parity-time (PT)-symmetric systems have received considerable research attention due to their various possible applications in optical devices. At EPs, at least two eigenvalues as well as their eigenvector coalesce. Recently, the effect of the eigenfrequency splitting on transfer function near EP was studied for an optical system consisting of two micro ring resonators, which led to complex splitting in PT-symmetric and real splitting in anti-PT-symmetric sensors. In present work, we propose a simple system of three coupled ring resonators to show real splitting in both PT-symmetric and anti-PT-symmetric parameter domains by exploiting higher-order EPs. We indirectly couple two rings with equal amount of gain and loss via an intermediate neutral ring. This system is then tested for refractive index (RI) sensing by modulating the cladding index and we numerically show a huge enhancement in sensitivity as compared to those reported in previous studies of micro ring resonators. Importantly, the enhancement is found to be of the order of $10^8$. Further, we have found that the order of EP can be tuned by perturbating the cladding. The outcomes of this study may set up a wide range of applications in non-Hermitian triplet cavity systems.

## 1. Introduction

The fundamental sensing mechanisms exploited in integrated optical devices are refractive index (RI) sensing, absorption sensing and fluorescence sensing. Among all, RI sensing is easy to implement and the most employed mechanism due to its label-free nature (i.e., additional chemical labels and markers are not required), high sensitivity, real-time monitoring capabilities and compatibility with optical fibers. It can be widely used in biomedical diagnostics, environmental monitoring, pharmaceutical development, food and beverage industry, and nanoparticle characterization [1]. During the last couple of decades, optical sensors with high sensitivity have been researched and developed extensively. Optical micro-ring resonators are the fundamental components of photonic integrated circuits [2]. Due to their compact size and high-quality factor, these structures have found extensive use as on-chip sensors in various configurations. In such systems, any perturbation introduced to the cavity can be detected by observing the consequent shift in the resonance frequencies [3, 4]. However, when we deal with micro-objects, sensitivity of the system decreases in a straightforward manner [5]. Physics of non-Hermitian exceptional points (EPs) of parity-time (PT)-symmetry and anti-PT-symmetry can provide the enhancement in sensitivity in such cases [5, 6].

Optical microcavities are the most suitable contenders for both micro and nano scale sensing applications, which support degenerate resonance frequencies. They work as a basic detecting element when a small amount of perturbation is introduced. This lifts the degeneracy and at the same time frequency splitting occurs [7]. The conventional degeneracies in microcavities are called diabolic points, which are linear reactions to any external perturbations. On the other hand, the degeneracies in PT-symmetric systems are called the EPs. At EPs, eigenvalues of the system collapse, and corresponding eigenvectors become parallel, leading to an abrupt change in dimensionality of the eigen-space. As a result, when an external perturbation is applied at this transition point, it not only lifts the degeneracy of the eigenvalues but also restores the system to its original dimensionality. In such cases, unlike in typical microcavities, the system responds to the square root of perturbation [8]. Due to extreme sensitivity of EPs, they have gathered significant research attention among photonic device research fraternities [9].

In recent years, PT-symmetric systems have led to a series of novel research outcomes due to the exotic property of exhibiting entirely real spectra even with non-Hermitian Hamiltonians. These systems show a phase transition at EP and beyond that the eigenvalues become complex [10-18]. This transition point (EP) may be utilized as an optimal condition for sensing applications [15, 19].

A system is called PT-symmetric if its Hamiltonian $\hat{H}$ commutes with $\hat{P}\hat{T}$ operator i.e., $[\hat{P}\hat{T}, \hat{H}] = 0$. On the other hand, in an anti-PT-symmetric system, the Hamiltonian anti-commutes with $\hat{P}\hat{T}$ operator that is $\{\hat{P}\hat{T}, \hat{H}\} = 0$. The parity operator $\hat{P}$ flips both the position and the momentum operators, i.e., $\hat{x} \to -\hat{x}$, $\hat{p} \to -\hat{p}$ while time reversal operator $\hat{T}$ inverts the flow of time which changes $\hat{p} \to -\hat{p}$, $\hat{\imath} \to -\hat{\imath}$ [10-18]. Usually, PT-symmetric systems are perceived via an active cavity and a passive cavity having identical resonant frequency with equal amount of amplification (gain) and absorption (loss) coefficients [19-24].

Several theoretical and experimental research have been done in the field of optical gyroscopes for enhancing sensitivity. The sensitivity in such systems is enhanced using different methods such as utilizing slow light effect, employing high quality factor, incorporating nano dimensional structures, fraction coating on ring surface, using multiple resonator configuration and using EPs of PT-symmetry [25, 26]. In addition, micro-cavity configurations have been studied in PT-symmetric isolators, and pumped lasers [14, 27].

Moreover, in recent years, anti-PT-symmetry has also been explored in many optical systems for sensing applications. Real eigen frequency splitting has been realized in a coupled anti-PT-symmetric sensor at micro and nanoscales [28, 29].

So far, both PT-symmetry and anti-PT-symmetry are utilized to strengthen the sensing characteristics of different types of optical sensing probes. It has been observed that most of the times PT-symmetry leads to complex splitting of the eigenfrequencies. On the other hand, anti-PT-symmetry yields real splitting. Recently, studies have been done for two coupled PT-symmetric micro cavities near EP for different sensing applications such as RI sensing, absorption sensing, and rotational sensing [5, 30].

In the present work, we elaborate RI sensing near EP in three coupled micro ring resonators in both PT-symmetric and anti-PT-symmetric domains. Importantly, here we propose a topology with real splitting in both PT-symmetric as well as anti-PT-symmetric domains. The three coupled micro cavities are shown schematically in fig. 1. This system is PT-symmetric in view of fact that under parity reflection operator the out rings get interchanged and under time reversal operator loss becomes gain and vice-versa.

The paper is arranged as follows- in Section 2, we present the theoretical background of PT-symmetry and anti-PT-symmetry for three-coupled micro cavities. Resonant shift induced perturbation approach is discussed in Section 3. In Section 4, numerical results for frequency splitting along with EPs order tunability and RI sensing characteristics of the optical system under various conditions have been deliberated. The results of the study are discussed and summarized in Section 5.

## 2. Theoretical background

To design RI sensor, we consider three coupled micro cavities. Any direct coupling between first and last rings has been neglected. Ring 1 and bus waveguide at the input end and ring 3 and the bus waveguide at output end of the system are coupled directly, as shown in Fig. 1. The red ring is an active resonator which can be made from Er3+ doped silica, grey ring has no gain/loss (neutral), and the green ring is a passive (no-gain-medium) resonator which can be fabricated using silica without any dopants [14, 31-33].

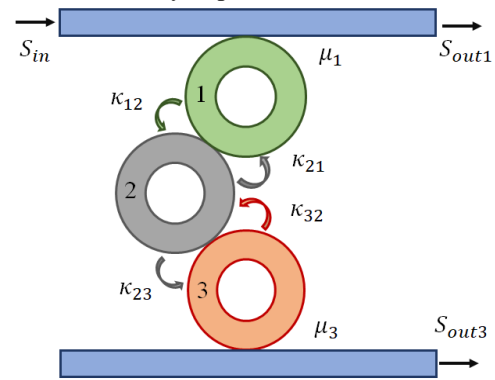

**Fig.1.** The schematic diagram of three coupled ring resonators with input and output bus waveguides.

### A. PT-Symmetric system

The necessary condition for a quantum system to be PT-symmetric is that the complex potential V of the system must follow $V(r) = V^*(-r)$. In the optical diffraction equation, the complex refractive index distribution is analogous to potential in quantum Schrödinger equation (which would result in necessary condition for PT-symmetric optical systems i.e., $n(r) = n^*(-r)$). Hence, in PT-symmetric optical system, the real component of the refractive index is symmetric in nature whereas the imaginary component is asymmetric and is responsible for gain/loss [21-24].

The unperturbed coupled system shown in Fig.1 can be described using the following equations [27, 30, 34],

$$\frac{db_1}{dt} = -i\omega_1 b_1 - \gamma_1 b_1 - i\kappa_{21} b_2 - \mu_1 S_{in}, \quad (1)$$

$$\frac{db_2}{dt} = -i\omega_2 b_2 - \gamma_2 b_2 - i\kappa_{12} b_1 - i\kappa_{32} b_3, \quad (2)$$

$$\frac{db_3}{dt} = -i\omega_3 b_3 - \gamma_3 b_3 - i\kappa_{23} b_2, \quad (3)$$

where $b_i$ ($i = 1,2,3$) is the normalized amplitude of CW propagating in $i^{th}$ micro ring, and $|b_i|^2$ is the energy stored in $i^{th}$ ring, $S_{in}$ represents normalized input amplitude, $\omega_i$ is the angular resonant frequency of $i^{th}$ unperturbed ring, $\gamma_i$ is the gain (if positive, neutral if zero and loss if negative) in $i^{th}$ ring, $\kappa_{ij}(\kappa_{ji})$ is the coupling strength between the $i^{th}$ ($j^{th}$) and the $j^{th}$ ($i^{th}$) rings, $\mu_{1,3}$ is the mutual coupling between the bus waveguide and the first (third) ring.

Output-input relation for this system can be expressed as [5]

$$S_{out1} = S_{in} - i\mu_1 b_1, \quad (4)$$
$$S_{out3} = -i\mu_3 b_3, \quad (5)$$

where $S_{out1}$ is the output amplitude of the field at the end of the top bus waveguide, while $S_{out3}$ is the field amplitude at the output end of bottom bus waveguide.

The Hamiltonian matrix of such a system can be derived from eqns. (1)-(3) and is expressed as:

$$H = \begin{bmatrix} \omega_1 - i\gamma_1 & \kappa_{21} & 0 \\ \kappa_{12} & \omega_2 - i\gamma_2 & \kappa_{32} \\ 0 & \kappa_{23} & \omega_3 - i\gamma_3 \end{bmatrix}. \quad (6)$$

As discussed earlier, for this device to be PT-symmetric, the commutation relation $[\widehat{PT}, \widehat{H}] = 0$ should hold, which gives $\omega_1 = \omega_3 = \omega_0$, $\gamma_1 = -\gamma_3 = \gamma_0$, $\gamma_2 = 0$, $\kappa_{12} = \kappa_{32}^*$, and $\kappa_{23} = \kappa_{21}^*$.

For simplicity, we have considered reciprocal coupling in first (third) and second ring i.e., $\kappa_{12} = \kappa_{21}$, and $\kappa_{23} = \kappa_{32}$ and the coupling coefficient of such system would be real in nature [5]. Also, we have assumed $\omega_2 = \omega_0$, and $\kappa_{12} = \kappa_{23} = \kappa_c$ in case of identical rings.

For the above discussed system, we write three supermodes, which we designate as central ($c$), upper ($u$), and lower ($l$) modes, respectively. The upper (lower) mode arises due to constructive interference (in the middle cavity) between two coupled two ring systems with symmetric (asymmetric) modes, having same (zero) phase for the fields in two side cavities (1 & 3) while central mode exhibits dark state in cavity 2 as shown in Fig.2. Here, '+' sign in the rings represents the symmetric mode while '-' sign in the rings denotes the antisymmetric mode [34-36].

Consequently, the resonant eigenfrequencies of the PT-symmetric system can be evaluated as

$$\omega_{PT_{c,u,l}} = \omega_0, \omega_0 \pm \sqrt{2\kappa_c^2 - \gamma_0^2}. \quad (7)$$

$\omega_{PT} = \omega_0$ for central frequency ($\omega_{PT_c}$), $\omega_{PT} > \omega_0$ for upper frequency ($\omega_{PT_u}$), and $\omega_{PT} < \omega_0$ for lower frequency ($\omega_{PT_l}$) respectively. The point of PT-symmetry breaking (called as EP) is reached when the expression in square root in eqn. (7) vanishes, i.e.,

$$2\kappa_c^2 - \gamma_0^2 = 0. \quad (8)$$

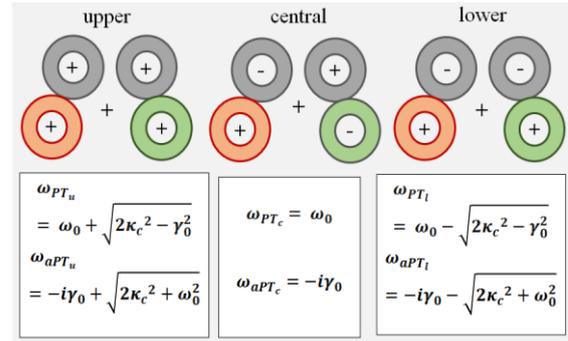

**Fig.2.** The superposition between modes of two coupled two rings for upper, central, and lower eigenmodes of three ring resonators having gain in green cavity, equal amount of loss in red ring and grey cavity is neutral.

### B. Anti-PT-Symmetric system

For system to be anti-PT-symmetric, Hamiltonian of the device should obey the anti-commutation relation i.e., $\{\widehat{PT}, \widehat{H}\} = 0$ which yields $\omega_1 = -\omega_3 = \omega_0$, $\omega_2 = 0$, $\gamma_1 = \gamma_3 = \gamma_0$, $\kappa_{12} = -\kappa_{32}^*$, and $\kappa_{23} = -\kappa_{21}^*$. Practically, negative resonant frequency does not make any sense. Hence, we may use quasi-anti-PT-symmetric system ($|\omega_1| \neq |\omega_2|$) [5]. As we know, zero resonant frequency is not possible for a resonator to be functional. So, we would choose it to be close to zero (i.e., $\omega_2 \to 0$).

In this case too we consider reciprocal system i.e., $\kappa_{12} = \kappa_{21}$, and $\kappa_{23} = \kappa_{32}$. We would like to notice here that the coupling coefficient would be imaginary for anti-PT-symmetric system which can be achieved using indirect coupling mechanism between resonators [20, 29, 30].

The resulting eigenfrequencies for anti-PT-symmetric system are given as

$$\omega_{aPT_{c,u,l}} = -i\gamma_0, -i\gamma_0 \pm \sqrt{2\kappa_c^2 + \omega_0^2}. \quad (9)$$

Again, for anti-PT-symmetric system EP can be found by equating square root term in eqn. (9) to zero, i.e.,

$$2\kappa_c^2 + \omega_0^2 = 0. \quad (10)$$

Table 1 summarizes the PT-symmetry and anti-PT-symmetry conditions in triple coupled micro ring resonators.

**Table 1. PT-symmetric and anti-PT-symmetric coupled resonator system.**

| Resonances | Gain/Loss | Coupling | Symmetry |
|---|---|---|---|
| $\omega_1 = \omega_3$, $\omega_2 \neq 0$ | $\gamma_3 = -\gamma_1$, $\gamma_2 = 0$ | $\kappa_{23} = \kappa_{12}^*$ | PT |
| $\omega_1 = \omega_3$, $\omega_2 \neq 0$ | $|\gamma_3| \neq |\gamma_1|$, $\gamma_2 = 0$ | $\kappa_{23} = \kappa_{12}^*$ | Quasi-PT |
| $\omega_1 = -\omega_3$, $\omega_2 = 0$ | $\gamma_3 = \gamma_1$, $\gamma_2 \neq 0$ | $\kappa_{23} = -\kappa_{12}^*$ | Anti-PT |
| $\|\omega_3\| \neq \|\omega_1\|$, $\omega_2 \to 0$ | $\gamma_3 = \gamma_1$, $\gamma_2 \neq 0$ | $\kappa_{23} = -\kappa_{12}{}^*$ | Quasi-Anti-PT |

## 3. Perturbation Approach

Phase transition (EP) is reached in both types of proposed configurations (PT- and anti-PT-symmetric) if the square root terms in eqns. (7) and (9) vanished. Since the applied perturbation induced frequency splitting makes the square root term nonzero and this ultimately results in enhancement in sensitivity in such sensing probe. We will use this concept in RI sensing. Here, we will see the effect of perturbation on behavior of output transfer function (normalized transmitted power at the end of the bottom waveguide), and eigenfrequencies for the coupled triple microcavities. Accordingly, we can design a methodical RI sensor with good enough sensitivity.

For RI sensing, as mentioned above, we will use resonance perturbation approach. This can be done in many ways, for example, we may expose one or more rings of the triple ring system to external resonant shift induced perturbation at a time. These different combinations of resonance perturbation will alter the output function, and accordingly the eigenfrequency spectra, which would result in enhanced sensitivity.

To design RI sensor, the conventional method has been applied in which refractive index is changed by exposing cladding index of the ring to external perturbation. This changes the cavity resonance wavelength $\lambda_R$ which is defined as

$$\lambda_R = \frac{n_e L}{s}, \quad (11)$$

where $s$ is the resonance order, $n_e$ is the effective index, and $L$ is the cavity (ring) length. Thus, a small variation in the refractive index of the cladding $(n_{cl})$ of a resonant cavity alters the effective index-

$$\Delta n_e = \frac{\partial n_e}{\partial n_{cl}} \Delta n_{cl}, \quad (12)$$

and

$$\Delta \lambda_R = \frac{\Delta n_e L}{s}, \quad (13)$$

which gives

$$\Delta \lambda_R = \frac{L}{s}\frac{\partial n_e}{\partial n_{cl}} \Delta n_{cl}, \quad (14)$$

and perturbation

$$\frac{\delta_\omega}{\omega_R} = -\frac{\Delta \lambda_R}{\lambda_R}, \quad (15)$$

Hence, resonance perturbation becomes:

$$\delta_\omega = -\frac{1}{n_e}\frac{\partial n_e}{\partial n_{cl}} \Delta n_{cl} \; \omega_R. \quad (16)$$

Consequently, the ring-resonator would encounter a change in its resonant frequency by $\delta_\omega$. A small alteration in cladding index $\Delta n_{cl}$ can lead to such a small resonance perturbation that may not be detected by usual devices (e.g., optical spectrum analyzer) [5]. PT-symmetric and anti-PT-symmetric triple coupled ring resonators operating at their EPs serve this purpose. Resonant perturbation $\delta_\omega$ in different rings would lead to a different output behavior.

Let us consider a case wherein the resonant shift induced perturbation $\delta_\omega$ is applied to first ring only $(\omega_1 \to \omega_1' + \delta_\omega)$, then the eigenfrequencies of the perturbed system are the roots of the characteristic equation as expressed as

$$\Omega(\omega) = (\omega_1' + \delta_\omega - i\gamma_1 - \omega)(\omega_2 - i\gamma_2 - \omega)(\omega_3 - i\gamma_3 - \omega) - \kappa_{23}\kappa_{32}(\omega_1' + \delta_\omega - i\gamma_1 - \omega) - \kappa_{21}\kappa_{12}(\omega_3 - i\gamma_3 - \omega) = 0. \quad (17)$$

Solving the above equation gives the eigenfrequencies $\omega_{PT_{c,u,l}}$ (central, upper, lower frequency), and $\omega_{aPT_{c,u,l}}$ for PT- and anti-PT-symmetric systems respectively.

Similarly, we can introduce perturbation in either second or third rings, or first and third rings simultaneously and then analyze the eigenfrequency spectra or energy spectra.

Also, we examine the behavior of the output transfer function $\left|\frac{S_{out3}}{S_{in}}\right|^2$ for all the above proposed perturbation configurations. Here, a broadband source can be used as input and a photoreceiver at the output to note the output reading.

## 4. Results and Discussion

In this section we focus on resonance perturbation induced effects on the sensing characteristics of the proposed three coupled ring resonator device. Simulation parameters are listed below: **(a)** for PT-symmetric case- $\omega_0 = 0$, $\mu_1 = \mu_3 = 1 \times 10^6 \; (rad/s)$, $\gamma_0 = 14 \times 10^6 \; (rad/s)$, $\kappa_{12} = \kappa_{23} = 14 \times 10^6 \; (rad/s)$, and **(b)** for anti-PT-symmetric case- $\omega_0 = 14 \times 10^6 \; (rad/s)$, $\mu_1 = \mu_3 = 1 \times 10^6 \; (rad/s)$, $\gamma_0 = 2 \times 10^6 \; (rad/s)$, $\kappa_{12} = -\kappa_{23} = 14\,i \times 10^6 \; (rad/s)$ [34].

### A. Effect of energy distribution

Since, the suggested device is linear and time invariant, the following condition is verified for the excitation angular frequency $\omega$ of the monochromatic source:

$$\frac{db_{1,2,3}}{dt} = i\omega b_{1,2,3}. \quad (18)$$

Substituting the above expression (18) in eqns. (1)-(3), we find:

$$\frac{b_3}{S_{in}} = \frac{-\kappa_{32}\kappa_{21}\mu_1}{\begin{bmatrix}[i(\omega-\omega_1)-\gamma_1][i(\omega-\omega_2)-\gamma_2]\,[i(\omega-\omega_3)-\gamma_3]+\kappa_{32}\kappa_{23}\\ [i(\omega-\omega_1)-\gamma_1]+\kappa_{12}\kappa_{21}[i(\omega-\omega_3)-\gamma_3]\end{bmatrix}}, \quad (19)$$

and the normalized output transfer function at the lower waveguide:

$$\frac{S_{out3}}{S_{in}} =$$

$$\frac{i\kappa_{32}\kappa_{23}\mu_1\mu_3}{\begin{bmatrix}[i(\omega-\omega_1)-\gamma_1][i(\omega-\omega_2)-\gamma_2][i(\omega-\omega_3)-\gamma_3]+\kappa_{32}\kappa_{23} \\ [i(\omega-\omega_1)-\gamma_1]+\kappa_{12}\kappa_{21}[i(\omega-\omega_3)-\gamma_3]\end{bmatrix}}. \quad (20)$$

To demonstrate the behavior of output for different sort of resonance perturbation approaches in PT-symmetric and anti-PT-symmetric systems, the normalized transmitted power at the output end of waveguide $\left|\frac{S_{out3}}{S_{in}}\right|^2$ is plotted in Figs.3-6.

*Case I. Perturbation in first ring*
When only the first ring is perturbed and others are kept isolated, both PT- and anti-PT-symmetric systems exhibit real eigenfrequency splitting and accordingly output power manifests actual splitting, which is shown in Fig.3.

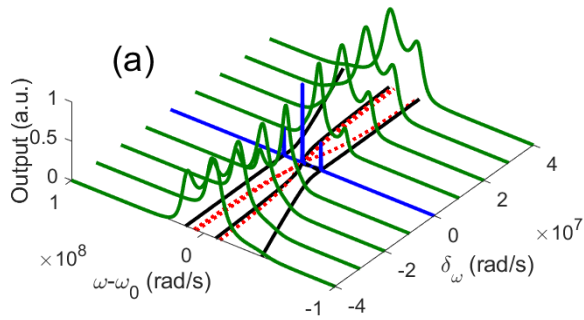
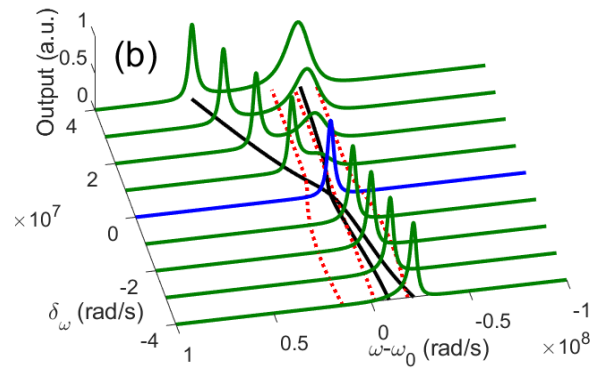

**Fig.3.** Evolution of eigenfrequency splitting and normalized output with resonant shift induced perturbation for (a) PT-symmetric, and (b) anti-PT-symmetric devices for the case when only ring 1 is perturbed (i.e., $\omega_1 = \omega_1' + \delta_\omega$). Also, real and imaginary parts of the eigenvalues are plotted with solid black lines and dotted red lines, respectively. Blue solid curve shows output for unperturbed system.

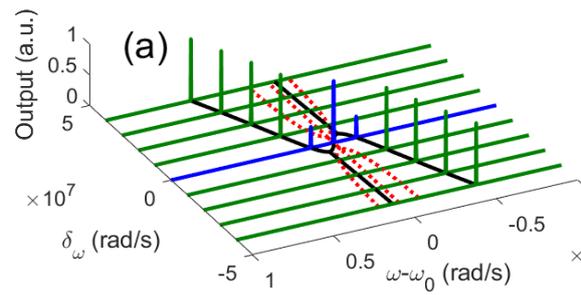
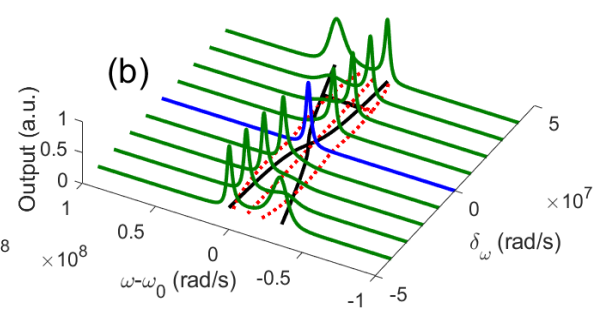

**Fig.4.** Evolution of eigenfrequency splitting with resonance perturbation for (a) PT-symmetric, and (b) anti-PT-symmetric devices for the case when only second ring is perturbed ($\omega_2 = \omega_2' + \delta_\omega$). Real and imaginary components of eigenfrequencies are plotted with solid black lines and dotted red lines, respectively. Blue curve represents energy distribution for unperturbed devices.

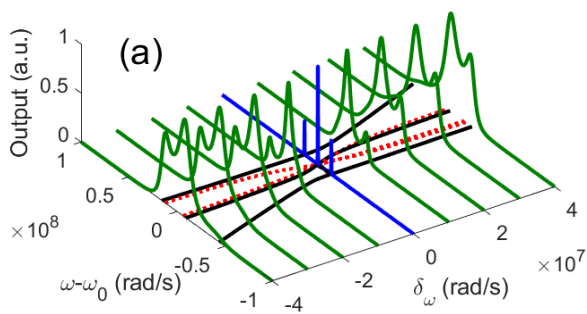
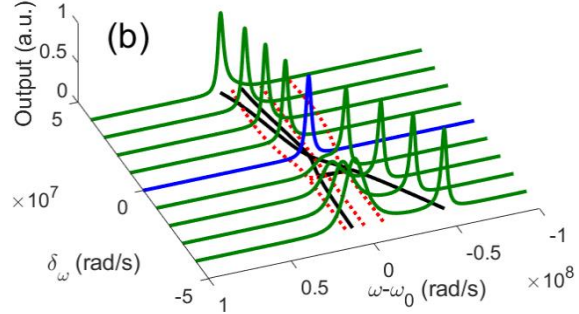

**Fig.5.** Eigenfrequency splitting along with output transmittance at bottom bus waveguide vs resonant shift induced perturbation for (a) PT-symmetric, and (b) anti-PT-symmetric devices for the case when only third ring is perturbed ($\omega_3 = \omega_3' + \delta_\omega$). Black lines in 2D plot denote the real frequencies and red dotted lines are imaginary frequencies. Blue curve represents normalized output at bottom bus waveguide for unperturbed system.

The PT-symmetric system provides splitting for both negative and positive resonant shift induced perturbations. The output of PT-symmetric system with positive and negative perturbations are inverted mirror image of each other as depicted in Figs.3(a)-6(a). On the other hand, anti-PT-symmetric system shows asymmetric and wider splitting than that in PT-symmetric system (see Fig. 3(b)). We would like to notice that in all 3D figures, the black curves represent the real eigenfrequencies of the coupled system and the red curves represent imaginary eigenfrequency components. The splitting in real components is an indicative of the splitting in normalized output power.

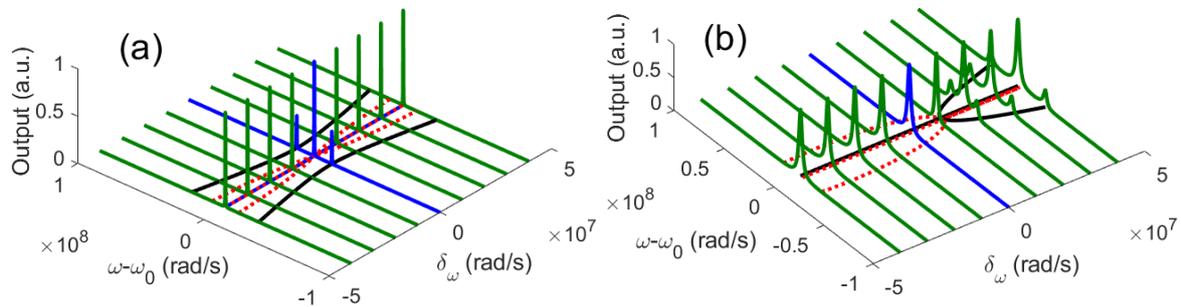

**Fig.6.** Evolution of energy distribution and eigenfrequency splitting with perturbed resonant frequency for (a) PT-symmetric, and (b) anti-PT-symmetric resonator when both first and third rings are perturbed simultaneously i.e., $\omega_1 = \omega_1' + \delta_\omega/2$, and $\omega_3 = \omega_3' - \delta_\omega/2$. Black and red solid lines in base plot denote the real and imaginary components, respectively. Blue line in 3D plot shows energy distribution for unperturbed system.

*Case II. Perturbation in second ring*

In this case, we perturb only the middle (second) ring of the device. The corresponding outputs are plotted in Fig.4. The splitting is observed in PT-symmetric system, but very little amount of energy is left in two splitted peaks (they are, therefore, very small in Fig.4 (a)) while a huge amount of energy resides in third sharp peak. Here in PT-symmetric system, two of three real eigenvalues collapse (black lines in Fig.4(a)). However, this behaviour of eigenfrequency splitting is prevalent for all resonant perturbation cases in anti-PT-symmetric system (Fig.3(b)-6(b)). Notice that the anti-PT-symmetric case shows the splitting with reciprocal behavior towards the sign of perturbation term as mentioned previously for PT-system as depicted Fig.4(b). We also note that the splitting in PT-symmetric structure is relatively large.

*Case III. Perturbation in third ring*

In this case, we introduce a small amount of perturbation in the third ring of the system. The output transfer function as well as eigenfrequency spectra in this configuration reciprocates all the results of case I (see Fig.5). This nature arises due to exchanged gain/loss parameters (equal and opposite $\gamma$ values) in first and third ring.

*Case IV. Perturbation in first and third rings simultaneously*

Next, we expose both the first and third rings simultaneously to external perturbation. As a result, the energy gets confined at one eigenfrequency component in PT-symmetric system whereas in anti-PT-symmetric system triple splitting (three eigenfrequencies) of energy is observed as shown in Fig.6.

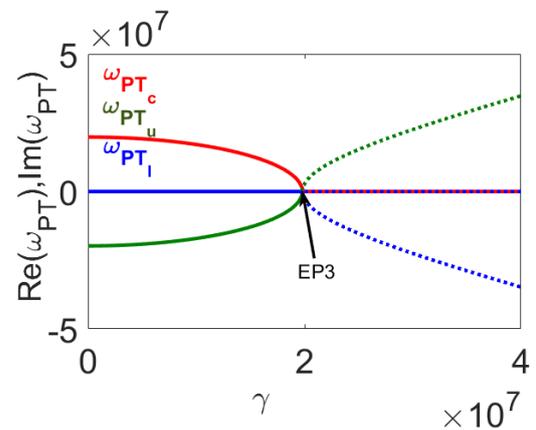

**Fig.7.** Eigenvalues (red color for central, blue for lower, and green for upper eigenmodes) are plotted as a function of the gain/loss parameter $\gamma$ for PT-symmetric system (solid lines represents real component while dotted lines show the imaginary components of eigenfrequencies) for unperturbed system (i.e., $\delta_\omega = 0$).

Notice that this triple splitting in anti-PT-symmetric system is obtained for positive resonance perturbation only while for negative values, all three eigenfrequencies merge into one eigenfrequency, which appears as single peak as shown in Fig.6(b). In Fig.6(a), the gain and loss terms in first and third ring balances in such a way that most of the energy gets confined with one eigenfrequency mode and others have negligible amount of energy, while in anti-PT-symmetric system, eigenfrequency splitting is either real or purely imaginary (depending on the sign of perturbation) as shown in Fig.6(b).

**B. Role of perturbation on EPs**

In addition to energy distribution analysis, we have also studied the effect of perturbation due to resonance shift, on modifying the order of EPs. For this, the eigenvalues are studied with variation in gain/loss parameter $\gamma$ for different values of the perturbation with different rings as shown in Figs.

7-9.

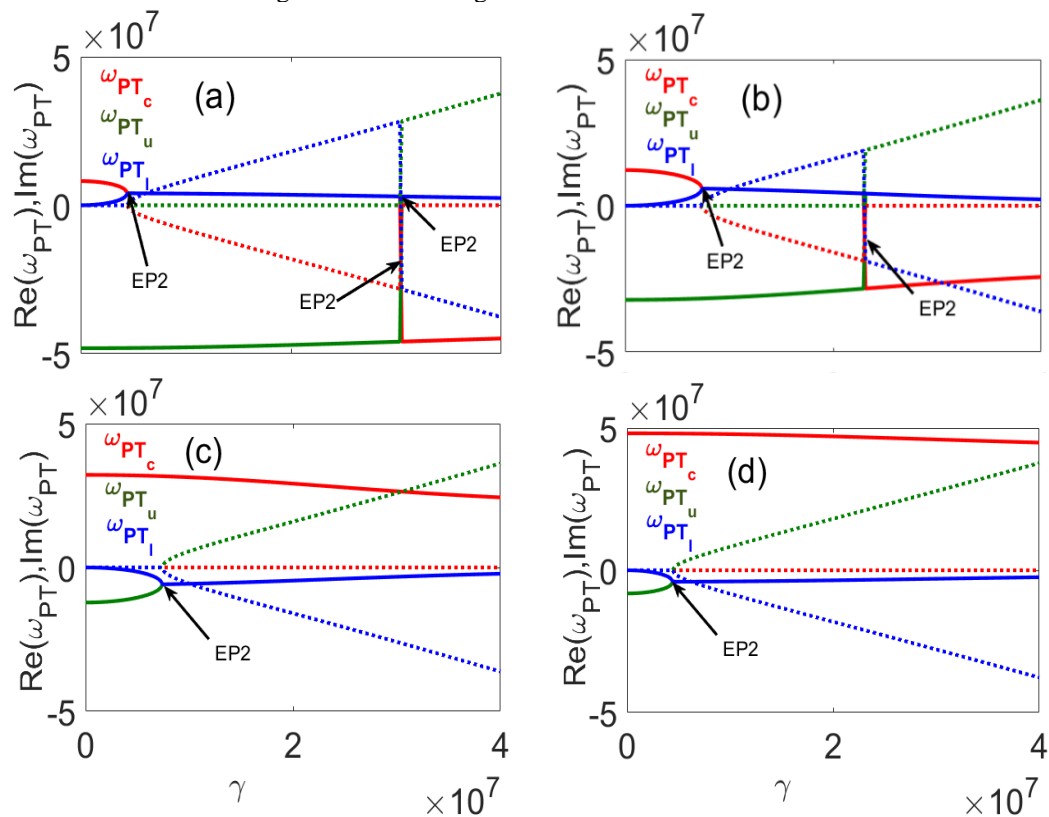

**Fig.8.** Eigenvalues (red for central, blue for lower, and green for upper eigenfrequencies) are plotted against the gain/loss parameter $\gamma$ for PT-symmetric system (solid lines for real parts while dotted lines for imaginary parts of frequencies) for (a) $\delta_\omega = -4 \times 10^7$, (b) $\delta_\omega = -2 \times 10^7$, (c) $\delta_\omega = 2 \times 10^7$, and (d) $\delta_\omega = 4 \times 10^7$ in second ring.

In unperturbed PT-symmetric system, real parts of all three eigenvalues collapse at a fixed value of gain/loss parameter and give rise to EP of order three (EP3) as shown in Fig.7.

Further, we observed that EP3 may be converted to EP2 by exposing any one of the three rings to an external perturbation as shown in Fig.8-9. This sudden change in the order of EP affects the eigenvalue spectra, which may lead to realization of highly sensitive sensing devices.

To demonstrate this, we added resonant perturbation to second ring and plotted eigenvalues against gain/loss parameter $\gamma$ [see Fig.8]. We observe that the eigenvalues are completely real below a particular value of $\gamma$. The EPs are shown by arrows in the figures. We would like to note here that for triple coupled ring system, when two of the three eigenvalues merge to one while the third eigenfrequency remains non-zero is called second order EP (EP2) and beyond this point, eigenvalues assume complex values.

We also observe that at negative value of resonant shift induced perturbation ($\delta_\omega = -4 \times 10^7$), central and lower eigenvalues coalesce (EP2) at low value of $\gamma$, and at larger values of $\gamma$ lower frequency component combines with upper one and central frequency gets separated with non-zero value. At even larger $\gamma$, this central frequency merges to upper frequency and lower frequency component gets separated (see Fig.8(a)). If we change perturbation value to $\delta_\omega = -2 \times 10^7$ then the positions of EPs get shifted as shown in Fig.8(b). For positive perturbation $\delta_\omega = 2 \times 10^7$, the upper and lower frequencies combine to form a single eigenvalue, and the central frequency remains nonzero as illustrated in Fig.8(c). Again, if we increase the perturbation to $\delta_\omega = 4 \times 10^7$, it shifts the EP towards lower value of $\gamma$ which can be clearly seen in Fig.8(d).

Shown in Fig.9(a-c) is the effects of the gain/loss parameter $\gamma$ variation on eigenfrequencies when only the first ring is exposed to the external perturbation. At $\delta_\omega = -2 \times 10^7$, real parts of upper and lower frequencies meet at low enough value of $\gamma$ and then bifurcate. Further increasing $\gamma$, upper and central real eigenvalues approach to zero and lower one remains non-zero as shown in Fig.9(a). If we take positive spectral shift due to perturbation ($\delta_\omega = 2 \times 10^7$), two EPs are observed at two nearby values of $\gamma$ for different eigenfrequencies as depicted in Fig.9(b). On further increasing resonance perturbation frequency ($\delta_\omega = 4 \times 10^7$) EPs shift towards higher gain/loss parameter (see Fig.9(c)).

Next, we will see how perturbation in third ring affects the eigenfrequency diagram (Fig.9(d)-9(f)). For $\delta_\omega = -2 \times 10^7$, we do not find any EP, but at larger value of $\gamma$, two eigenvalues try to approach their zero as plotted in Fig.9(d). Now for positive disturbance ($\delta_\omega = 2 \times 10^7$), we find three different configurations of EPs (see Fig.9(e)). Again, if we increase perturbation to $\delta_\omega = 4 \times 10^7$, shift in EPs is obtained as given in Fig.9(f).

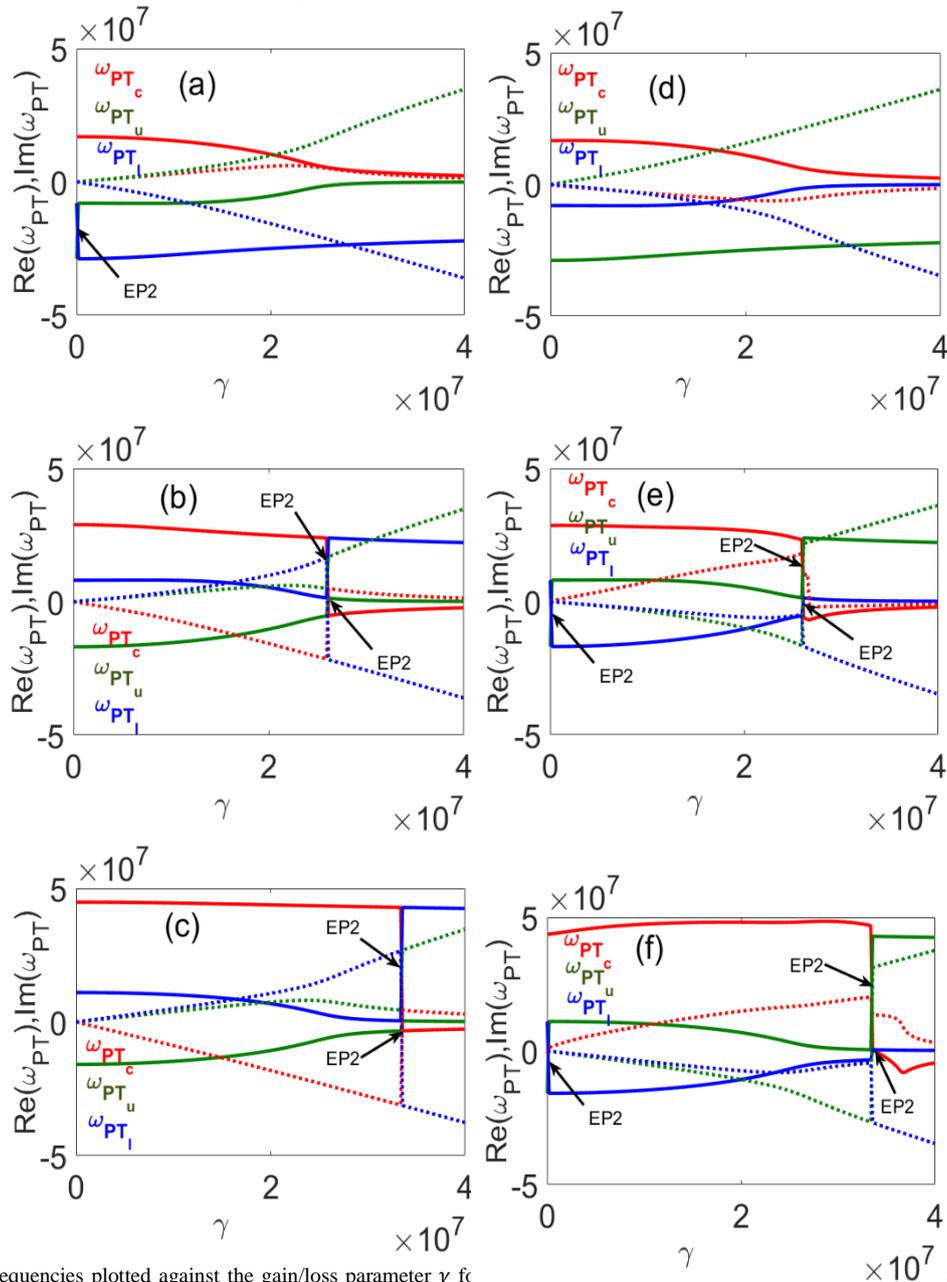

**Fig.9.** Eigenfrequencies plotted against the gain/loss parameter $\gamma$ for [...] when resonance perturbation is introduced in first ring and right column for the case when third ring is perturbed) (solid lines for real parts while dotted lines for imaginary parts of frequencies) (a, d) $\delta_\omega = -2 \times 10^7$, (b, e) $\delta_\omega = 2 \times 10^7$, and (c, f) $\delta_\omega = 4 \times 10^7$.

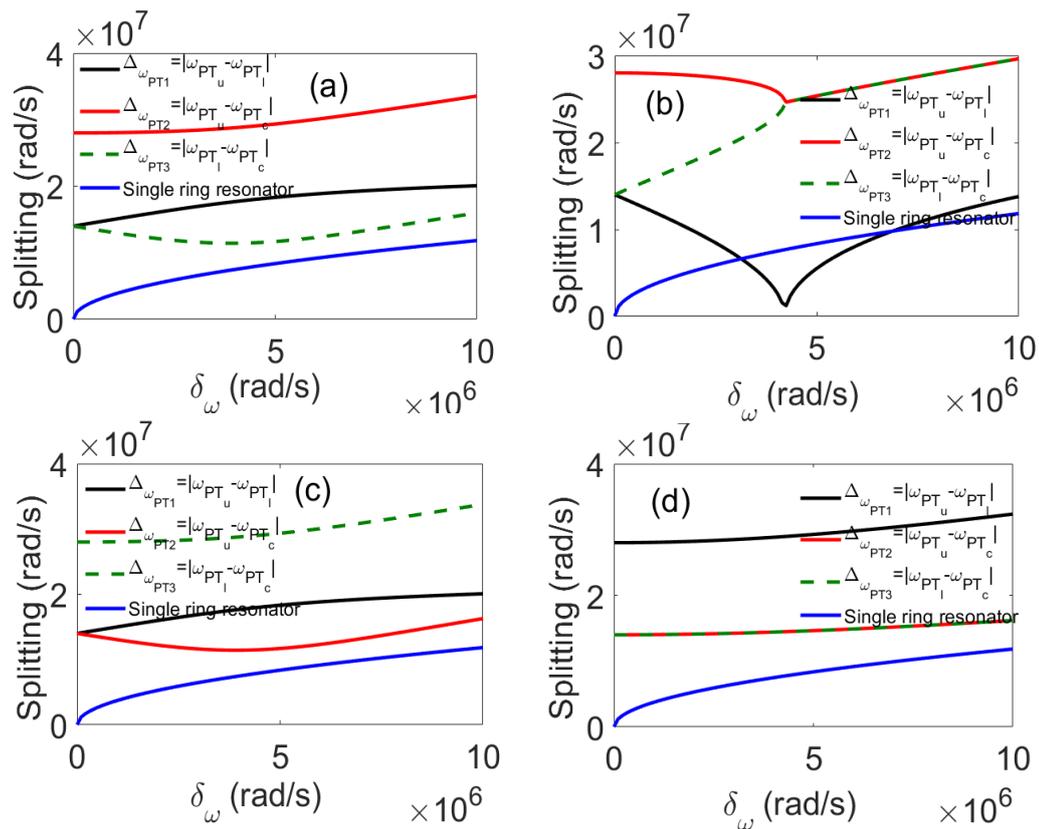

**Fig.10.** Spectral splitting at the exceptional point for PT-symmetric RI sensor when perturbation in (a) upper ring, (b) middle ring, (c) bottom ring, and (d) first and third rings simultaneously.

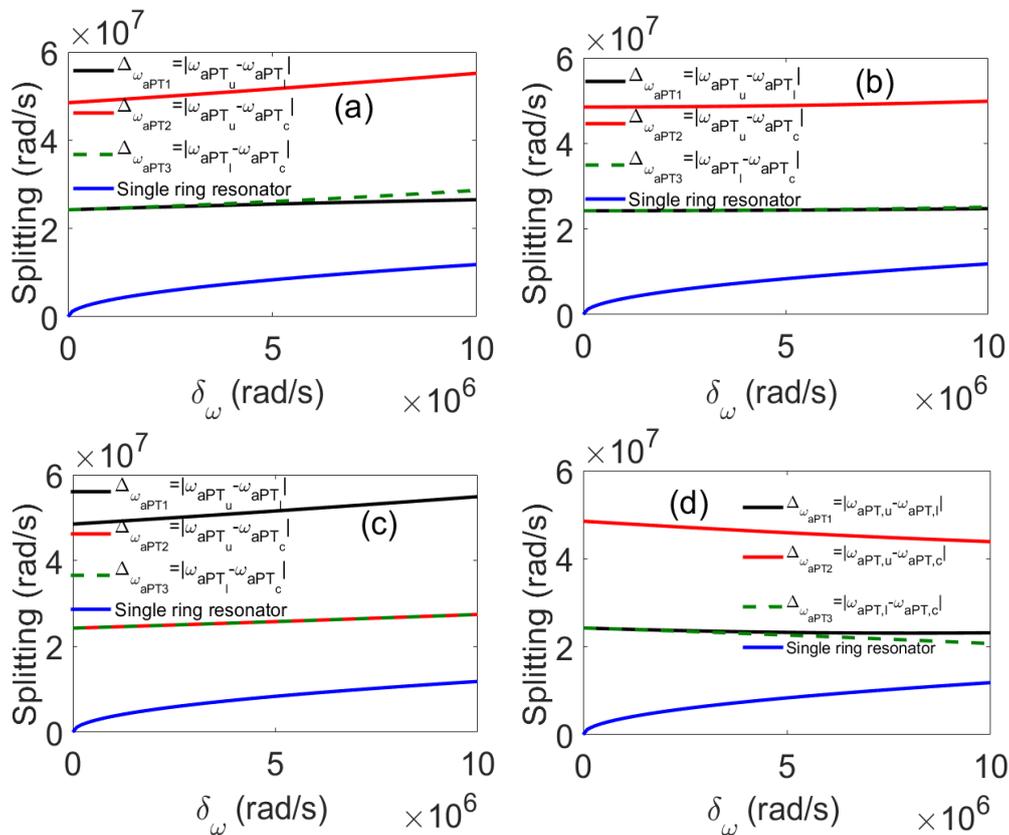

**Fig.11.** Spectral splitting near EP for different perturbation configuration for anti-PT-symmetric device with perturbation in (a) ring 1, (b) ring 2, (c) ring 3, and (d) ring 1, ring 3 simultaneously.

### C. Spectral splitting enhancement and sensing

Shown in Fig.10 and Fig.11 are plots for spectral splitting vs resonance perturbation frequency for different perturbed combinations.

Unlike conventional single ring resonator (blue solid line) based sensors, the splitting of triple coupled rings resonator does not linearly depend on resonant shift induced perturbation. It is directly proportional to the cube root of external perturbation due to resonance spectral shift.

Here, the sensitivity enhancement factor can be measured as $S = \Delta\omega(\delta_\omega)/\delta_\omega$, where $\delta_\omega$ is resonance perturbation and $\Delta\omega$ is the difference between two of the three (upper, lower, and central) eigenfrequency modes i.e., $\Delta\omega_{PT1} = |\omega_{PT_u} - \omega_{PT_l}|$, $\Delta\omega_{PT2} = |\omega_{PT_u} - \omega_{PT_c}|$, and $\Delta\omega_{PT3} = |\omega_{PT_l} - \omega_{PT_c}|$. Similarly, we can find $\Delta\omega$ and then sensitivity $S$ for anti-PT-symmetric case. Based on the above calculation, we noticed that the value of the sensitivity enhancement factor depends on coupling coefficient between micro rings.

For both PT- and anti-PT-symmetric systems, in all perturbed cases discussed in this paper, we achieved $10^8$ times enhanced in the maximum sensitivity compared to classical optical ring resonator sensor which can be calculated using above mentioned sensitivity enhancement factor $S$ (See supplementary). The sensitivity for single ring resonator can be calculated using $S = \kappa/\delta_\omega$, which has a maximum value of order of $10^2$ (assumed coupling $\kappa = 10^{-4}$ small enough to be near zero) [5]. Hence, we exploit EPs to get enhanced sensitivity. The superiority of this type of sensor is that they can be utilized to sense even a small amount of perturbation compared with that in a classical optical sensor.

### 5. Conclusion

In conclusion, we have studied an optical system consisting of three coupled rings with PT- and anti-PT-symmetric nature for RI sensing by exploiting EP to enhance the sensitivity of the suggested device. Different perturbed configurations have been explored. We observed better sensitivity ($10^8$ times enhancement) as compared to single resonator case which may be utilized in several photonics devices. On the other hand, it is noticed that even PT-symmetric system offers real eigenfrequency splitting along with anti-PT-symmetric system. Earlier, real eigenfrequency splitting was reported only in the case of anti-PT system. Moreover, we observed that the resonant shift-based perturbation approach may sensitively change the order of EPs in the above discussed system. Practically, achieving the ideal EP condition is quite challenging. Therefore, we conduct research on quasi-non-Hermitian systems to improve sensitivity based on EP. This approach has the potential to open up new possibilities for enhancing sensitivities that can be achieved with integrated photonic devices.

### Credit Authorship Contribution Statement

**P. Chaudhary:** Design; Methodology; Analytical calculations; Numerical simulations; Writing-original draft. **A. K. Mishra:** Conceptualization, Writing – review & editing, Supervision.

### Declaration of Competing Interest

The authors affirm that there are no identified competing financial interests or personal relationships that could have potentially influenced the findings presented in this paper.

### Data Availability

Data will be available on request to promote transparency and facilitate scientific collaboration.

### Acknowledgement

Priyanka Chaudhary acknowledges IIT Roorkee and the Ministry of Education (MOE), Government of India, for the research fellowship to undertake her Ph.D. studies.

### Appendix A. Supporting Information

Supplementary data for this article is available at the following link.